\documentstyle[preprint,aps]{revtex}

\tighten

\begin{document}
\title{Potential Effect: Aharonov-Bohm Effect of Simply Connected Region
\footnote{Los Alamos National Laboratory E-Print Archive, quant-ph/9506038
 (June 25, 1995)}}
\date{Revised June 25, 1995}
\author{J\`un L\'\i u}
\address{Physics Department, Rensselaer Polytechnic Institute,
Troy, NY 12180-3590 \footnote{{\it e-mail}: liuj3@rpi.edu}}
\maketitle

\vspace{0.2in}

\begin{abstract}
Starting from variations on the Aharonov-Bohm effect, we suggest that any
non-trivial electromagnetic potential, vector or scalar, can generate a
measurable effect on charged particle. The title experiment is a realizable
and clean example of this. This effect is not topological, and can be tested
by a diffraction experiment inside a large-enough elongated toroidal
solenoid. A wave-front theory based on four-momentum conservation is
introduced as interpretation.
\end{abstract}

\vspace{0.5in}
It took thirty-three years after the discovery of
Schr\"odinger equation for people to realize the existence of the
Aharonov-Bohm effect \cite{AB,Exp} (hereafter AB effect). It is still a
vital topic today \cite{AC,exp,Magni,AA}. In this paper, we shall study a
generalization of AB effect, the potential effect as described in abstract.
The discussion is focused on effects in simply connected region, which
obviously can not have any local field-flux. Among the published discussions
about this kind of effects,
it is generally agreed
that this kind of effects does not exist due to gauge invariance.
For example, \cite{referee1}. However, there are also opinions that this
effect is a trivial variation of AB effect and therefore there is no need to
check its existence \cite{referee2}. To my knowledge, it has never been
tested. My first goal here is to supply enough theoretical reason to
motivate the experimental test of this effect. I start with an intuitive
derivation in {\it b}, then I introduce a wave-front theory as a theoretical
consideration. Logically, the existence of potential effect implies the
existence of the AB effect, but not vice versa. The purpose of this paper is
to provide a physical connection in the opposite direction. I wish the
reader to understand from the very beginning that this is not a mathematical
derivation from any existing first-principle.

\paragraph{Is the original AB effect enough for testing the importance of
electromagnetic potential?}

Experimental tests of AB effect are still developing in order to provide
more convincing results \cite{Tonomura,exp,PT}. The major issue in
experimental design is to shield the field-flux into a small region to
ensure that it is an AB effect. This is however not easy mainly because the
typical wave-length of an electron beam is only 0.03\AA\ \cite[p.101]{PT}.
The construction has to be very small. In the dual version of AB effect, the
Aharonov-Casher effect \cite{AC}, the magnetic flux from the neutron is
basically not shielded. Also, impenetrable wall is a grey concept. The
incident partial wave-front is either reflected or absorbed, or both. This
complicates the interpretation and boundary condition, because the reflected
wave-front can carry information of the surface and then interferes with the
rest of the wave-front \cite{AS}. Another problem is the effect of returning
flux. In order to put this problem away, Peshkin, Talmi and Tassie \cite{PTT}
discussed a two-solenoids arrangement as shown in Fig.I(2). According to
their calculation at several angles, they concluded that the scattering of
electrons would be altered even when the returning flux was taken into
consideration. Exact solutions of the ideal double-flux scattering have been
attempted, but the results are so far not unique \cite{GQ}.
The situation was properly
commented by Aharonov and Bohm themselves \cite{AB}, and by Tonomura
\cite[p.53]{PT}, that none of the experiments could be regarded as an ideal
confirmation of the significance of the potentials in quantum mechanics.

The literature of the last thirty-six years indicates that the
connection between AB effect and quantum mechanics is so far less than
perfect. The problem is associated with the special geometry, topology and
boundary conditions of the AB effect.

\paragraph{Moving the field-flux away}

We start our discussion with Fig.I(3). Two flux-tubes of opposite directions
are put side-by-side in front of the double-slit like a dipole. The
influence of a single flux-tube, if any, must be cylindrical with respect to
its center. Therefore, the effects of two non-coaxial solenoids can not
possibly cancel each other out completely. There should be a remaining
higher-order effect for Fig.I(3). We now move the two flux-tube out, see
Fig.I(4). The geometry is changed, but based on the same argument, there
should be a similar higher-order effect. We estimate qualitatively what the
remaining effect should look like. The ${\bf A}$ potential for Fig.I(3) has
right-left symmetry. Therefore, the change of interference pattern should
also carry this symmetry, and the central maximum should not move. Also, we
can expect the change to be uniform since the potential is not periodic in
space. The possible change to the interference pattern is demonstrated in
Fig.II.

Effects from Fig.I(3) and Fig.I(4) are obviously small because the
first-order effect from single flux-tube is canceled out. There is a
similarity between these designs and the toroidal ferromagnet experiments by
Tonomura {\it et al.} \cite{Tonomura}. Their experiment is a good
inspiration, but the effect we wish to discuss, the change of wave-length,
is too small to be detected with the deposited toroidal ferromagnet. On the
other hand, although their experiment is conducted with an incident plane
wave, their phase-shift is measured by counting fringes in the shaded
interference region. Therefore, the effect they measured is still
effectively a multi-connected AB effect.

Fig.I(5) represents an experiment that is realizable and the region is
simply connected. A big-enough elongated toroidal solenoid is used to cover
the experimental region after the double-slit. Fig.I(5) can be viewed as the
result of moving the two flux-tube away while increasing the strength of the
flux. By doing so, we effectively eliminated the flux-tube in the
interference region. But is the above effect of Fig.I(3) and I(4) still
there?

We use another connection to the original AB effect through Fig.I(6). In
Fig.I(6), only one path is covered by the toroidal solenoid. The total flux
in the whole closed path is non-zero. We can see that, as a normal AB
effect, there is a phase-shift in the closed path, {\em or}, we can
equivalently say that the wave-front in the right-channel travels at a
different wave-length compared to the normal left-channel. Our goal is to
connect this picture(Fig.I(6)) to Fig.I(5) by cutting away the left-channel.
The wave-fronts in the left-channel and the right-channel are coherent.
However, we can argue that, due to causality, whatever happened to one
channel would not affect the {\em propagation behavior }of the other
channel. After cutting away the left-channel, we assume the wave-front in
the right-channel alone still travels at the changed wave-length. This
change of wave-length means a {\em potential effect} (see Fig.II) for
experiment of Fig.I(5). Experiments can be done in five variations, with the
toroidal solenoid covers (1) all except source, (2) region after double
slit, (3) double-slit and source, (4) double-slit only, (5) source only.
Whether it is positive or negative, to my knowledge, this experiment of
Fig.I(5) has never been reported.

So potential effect of Fig.I(5) is an effect of changing wave-length, while
AB effect of Fig.I(1) is an effect of shifting the central maximum.
Obviously, both effects are field-free.\vspace{0.5in}

\paragraph{AB effect revisited}

Ten years before Aharonov and Bohm \cite{AB}, Ehrenberg and Siday \cite{ES}
formulated electron optics with refractive index represented by the
potentials. Fig.I(1) was their Figure 3. Here, {\em we look for a more
general observable which could be named field-free electron optics, with AB
effect as a special case of it}. The issue turns out to be more conceptual
than numerical, to begin with. The AB effect is known to inspire discussion
about the importance of phase in quantum theory. Wu and Yang \cite{WY}
stated that phase-factor provides complete description of the classical
electromagnetic field. Aharonov and Anandan \cite{AA} discussed the problem
whether there is a preferred quantum gauge or not. They cure their preferred
gauge problem by including an acceleration dependent term in the Lagrangian.
The phase factor of a closed path
\begin{equation}
\label{1}u_r=exp\left( -\frac{ie}\hbar \oint A_\mu dx^\mu \right)
\end{equation}
is clearly gauge invariant. We need phase factor for open path here, so we
also need to discuss the gauge invariance problem.

There are variations on the interpretation of AB effect.
Zhu and Henneberger \cite
{ZH} introduced $\left( e/c\right) {\bf A}$ in a Coulomb gauge as the
electromagnetic momentum. Therefore, in their theory, effects of vector
potential look like effects of total momentum conservation. Another
inspiration comes from the interpretation of Rubio, Getino and Rojo \cite
{RGR}. They tried to explain the AB effect as a classical electromagnetic
effect. Two particles traveling on the right and left side of the solenoid,
respectively, acquire different acceleration due to an equation%
$$
\frac{\partial {\bf p}}{\partial t}=-q\frac{\partial {\bf A}}{\partial t}.
$$

Our interpretation of the AB effect, in the spirit of the previous
conceptual discussion in {\it b}, is:
$$
\begin{array}{c}
\text{Vector potential }{\bf A}\text{ causes change in local wave-vector,}
\\ \text{scalar potential }\phi \text{ causes change in local frequency.}
\end{array}
$$
Although the above statements sound familiar, it is different. The key word
is "local", which means, non-topological. The standard interpretation is
topological \cite{AB}. In the standard interpretation, the potential
generates a phase-factor which is not realized until partial waves of
different
channels meet each other with a gauge-invariant phase-difference. This
complex interpretation created a gap between quantum mechanics and the AB
effect. Our interpretation is actually closer to the original interpretation
of Ehrenberg and Siday \cite{ES}, which is sometimes called the classical
interpretation.

By our interpretation, {\em %
any non-trivial electromagnetic potential, }${\bf A}${\em \ or }$\phi ${\em %
, can cause a measurable effect on the charged particle}. In particular, it
support our previous analysis of the potential effect of Fig.I(5), with its
result shown in Fig.II.

\paragraph{A wave-front theory}

The quantum mechanics we have today is actually weak, not as effective
compared to Neutonian mechanics and special relativity. We are already
accustomed to its approximate solutions. No
matter it is solid state physics or particle physics, we always have to
depend on heavy modelings to solve problems. Usually, The modelings become
so complicated that they themselves become the real theories. We used to say
that everything is explained by quantum mechanics, but if we count all its
modelings, the size of this theory is huge and still growing.
We rarely have such problems
with Neutonian mechanics. Similar logic must have inspired legends like
Einstein, de Broglie, Schr\"odinger and so on. It is not the goal
of this particular paper
to establish a whole new theory, though. I merely want to develop a slightly
different theory, which
may eventually be resolved as a modeling over quantum mechanics, or may not.
The motivation is its simplicity, and its potential generality if the effect
of this paper is experimentally verified.

We of course expect AB effect to be part of a
larger quantum theory. Usually, the quantum mechanics is extended through
modeling to interpret effects like this. We are going to do the reverse. Let
us extend the above interpretation into a full theory of field-free
electromagnetic interaction:

\begin{enumerate}
\item  A particle with charge $q$ generated at point ${\bf r}_0,t_0$ travels
as a probability wave of amplitude $\psi $. It starts with $\nu _0=E_0/h$, $%
{\bf k}_0={\bf p}_0/\hbar $. Probability-amplitude relation is $\rho =|\psi
|^2$.

\item  A region filled with electromagnetic potential behaves as a media of
variable index, comparable to an optical media.
\begin{equation}
\label{2}\nu \left( {\bf r},t\right) =\nu _0+q\left( \phi \left( {\bf r}%
,t\right) -\phi \left( {\bf r}_0,t_0\right) \right) /h,
\end{equation}
\begin{equation}
\label{3}{\bf k}\left( {\bf r},t\right) ={\bf k}_0+q\left( {\bf A}\left(
{\bf r},t\right) -{\bf A}\left( {\bf r}_0,t_0\right) \right) /\hbar .
\end{equation}

\item  Any scattering problem can be solved by starting from a desired
wave-front, and then see how it evolves in the media of potential, using
methods similar to that of optics. Non-eigenstate of total four-momentum
must be first expanded into a sum of eigenstates. Result is the sum of
scattering results of its components.

\item  A bounded eigenstate is a state that scatters into itself.
\end{enumerate}

Conservation of energy on each point means a connection to the Klein-Gordon
equation (hereafter KG equation, with Dirac equation implied). The
connection is as follows:
$$
\begin{array}{c}
\left[ \nu \left(
{\bf r},t\right) h-q\phi \left( {\bf r},t\right) \right] ^2-\left[ {\bf k}%
\left( {\bf r},t\right) \hbar -q{\bf A}\left( {\bf r},t\right) \right]
^2=m^2 \\ =\left[ \nu \left( {\bf r}_0,t_0\right) h-q\phi \left( {\bf r}%
_0,t_0\right) \right] ^2-\left[ {\bf k}\left( {\bf r}_0,t_0\right) \hbar -q%
{\bf A}\left( {\bf r}_0,t_0\right) \right] ^2.
\end{array}
$$
This connection ensures the general effectiveness of this representation.
Quantum mechanics stated this way seems very suitable for computer
simulations of wave-front evolution. It also provides a natural
interpretation of the potential effect.

We can speculate on the resemblance between this theory and more standard
quantum mechanics of Schr\"odinger representation and path-integral
representation. It is basically the conservation of a four-momentum $k_\mu
\hbar -qA_\mu $. Klein-Gordon equation means the conservation of its module:
the mass. So there is a difference, but no conflict. We may need correction
terms for (\ref{2},\ref{3}) if we do not wish to expand initial states into
eigenstates of total momentum. We can expect the four-momentum to rotate in
the presence of field:%
$$
d\left( \hbar k^\alpha -qA^\alpha \right) =\frac qm\left( \partial ^\alpha
A^\beta -\partial ^\beta A^\alpha \right) \left( \hbar k^\beta -qA^\beta
\right) d\tau .
$$
The details of this complication shall be discussed separately. We return to
the field-free case.

\paragraph{Gauge invariance}

Many people do not believe the existence of potential effect in simply
connected region because they think that the result can always be ``gauged
out''. This is an oversimplification. The AB effect also ``disappears''
under certain gauge transformations \cite{BL}, despite the fact that it
never disappears in experiments.
When a theory gives gauge-dependent
prediction, it could be that the theory itself is incomplete.

In order to show the gauge invariance of our wave-front theory, we start
with a potential from a gauge transformation:
\begin{equation}
\label{A}{\bf A}\left( {\bf r},t\right) =\nabla S\left( {\bf r},t\right)
,\;\phi \left( {\bf r},t\right) =-\frac \partial {\partial t}S\left( {\bf r}%
,t\right) .
\end{equation}
By calculating the phase from (\ref{2},\ref{3}), we have
\begin{equation}
\label{psi}\psi =c\exp \left( \frac{iq}\hbar S\left( {\bf r},t\right)
\right) ,
\end{equation}
This is the known uncertainty to the wave-function from the gauge
invariance. It also arrives naturally in this approach. We can check that
the total four-momentum is gauge-invariant: for eq.(\ref{A},\ref{psi}), $%
\left( i\partial ^\mu -qA^\mu \right) \psi =0$.

We now proceed to consider our own problem: whether the wave-front theory
prediction of this potential effect is gauge-invariant or not.
we assume the source is at infinity and
toroidal solenoid is of finite length. We can set ${\bf A}$ to zero in the
interference region through a gauge transformation:%
$$
{\bf A}^{\prime }\left( 0\right) ={\bf A}\left( 0\right) +\nabla S\left(
0\right) =0.
$$
But as a result of the transformation, we also have
$$
S=-\int {\bf A}\left( 0\right) \cdot d{\bf r,\;}\Rightarrow {\bf A}^{\prime
}\left( \infty \right) =0+\nabla S=-{\bf A}\left( 0\right) .
$$
This transformation does not change the potential difference between the
source and the interference region, so according to (\ref{3}), it does not
change the prediction of potential effect of Fig.I(5).

According to this wave-front theory, the source also changes its phase
under a gauge transformation. This is the reason that our prediction is
gauge invariant.

We now proceed to discuss a more critical issue, the local expectation of $%
E^2-{\bf p}^2$. Let us assume that a particle is generated at point 0 and
then observed at point 1, and the people operating the generator/detector
have no knowledge of the local potentials. When they generate/detect a
particle of frequency $\nu $ and wave-length $\lambda $, they may conclude
that the mass is $\left( h\nu \right) ^2-\left( h/\lambda \right) ^2$, that
is, $m_0$ or $m_1$ defined as%
$$
m_0^2=\left( h\nu _0\right) ^2-\left( h/\lambda _0\right) ^2,\;m_1^2=\left(
h\nu _1\right) ^2-\left( h/\lambda _1\right) ^2.
$$
Notice that generally speaking,
\begin{equation}
\label{m}m\neq m_0\neq m_1,
\end{equation}
because $E^2-{\bf p}^2$ and $\left( E-q\phi \right) ^2-\left( {\bf p}-q{\bf
A%
}\right) ^2$ generally do not commute. This means $\left\langle
E\right\rangle _{area}^2-\left\langle {\bf p}\right\rangle _{area}^2$ is
area dependent for an eigenstate of KG equation. It becomes important in
this approach because here we seek solution from propagating wave-front.
People by the detector normally need to know potential difference between
point 1 and point 0 to determine its original mass from $\lambda _1$ and $%
\nu _1$.

Although there might not be any preferred quantum gauge \cite{AA}, there can
certainly be more convenient quantum gauges. We can reasonably choose the
gauge to let the new-born particle has zero field-momentum, so that $m_0=m$.
Following this line, it seems that we can measure the values of local
potentials with respect to the particle source in a field-free case. The
value and direction of ${\bf A}$ (with respect to the particle source) can
be measured by a device made after Fig.I(5), the fourth equation comes from
additional $m$ measurement by trajectory radius in a magnetic field and $%
h\nu $ measurement.

\paragraph{Is there an alternative to this prediction?}

We can think of other theoretical possibilities before the experiment. In
avoiding this problem under this wave-front picture, it is possible to have
an interpretation of AB effect which {\em denies} the existence of the
potential effect. For the free-of-field case: (1) Wave-front propagates with
wave-length $\lambda =2\pi /|{\bf k}_0|$ in any simply connected region. (2)
Multi-connected region is divided into a {\em minimum }number of simply
connected sub-region. (3) When wave-fronts from two different simply
connected sub-region meet and combine, their relative phase is calculated
{\em as if} the wave-fronts travelled through each with ${\bf k}={\bf k}_0+q%
{\bf A}/\hbar $. Notably, the Tonomura {\it et al.} experiments \cite
{Tonomura} can also be explained by this interpretation. This interpretation
puts all emphasis on geometry \cite{Berry}. It is not as natural as the
previous interpretation.

\paragraph{Conclusion and numerical prediction}

To conclude, we find experiment of type Fig.I(5) worthwhile. The double-slit
can be replaced by a single-slit, circular opening, grating, crystal
surface, or even a hologram. Any method sensitive to the wave-length can be
applied, since the region is simply connected. The scalar version of this
effect should be designed carefully so that no field is introduced. In an
ideal design similar to Fig.I(5), no change of interference should happen to
a changing V(t). The dual Aharonov-Casher \cite{AC} like potential effect is
also possible with neutron interferometry \cite{exp}, with result Fig.II.
Ideal potential effect of Fig.I(5) with a long-enough toroidal solenoid is:
\begin{equation}
\triangle \left( \frac 1\lambda \right) =\left( \text{thickness of toroidal
solenoid}\right) \times \frac{qB}h.
\end{equation}
For example, 0.01m and 0.01Tesla means about 7\% change to the wave-length
of a 0.03\AA\ electron. Effect on screen is basically the same as the result
of moving the screen forward/backward by the same percentage. At certain
value of B, the interference pattern would disappear.

This effect should have important theoretical consequence. (1) There will be
no need to test AB effect separately, it is implied. (2) The wave-front
theory mentioned here may become a major representation of quantum theory.
(3) Due to the simplicity and generality of this effect, it can have much
wider application compared to that of AB effect in condensed matter physics
and electron optics. A classical particle is not influenced by a curl-free
vector potential because $\delta \left( {\bf r}\right) $ contains equal
amount of all wave-vectors. So there is no conflict with classical
electrodynamics.

\vspace{0.2in}

\begin{center}
{\bf FIGURE CAPTIONS}
\end{center}

\begin{description}
\item[Fig. 1: ]  (1) AB effect. (2) Peshkin {\it et al.}'s discussion \cite
{PTT} about returning flux. (3-4) There should be a higher order effect
remaining, which can be magnified into: (5) The title effect of this letter.
(6) Cutting away the left channel should not affect the propagation property
of the right.

\item[Fig. 2: ]  Results of ideal double-slit.
\end{description}

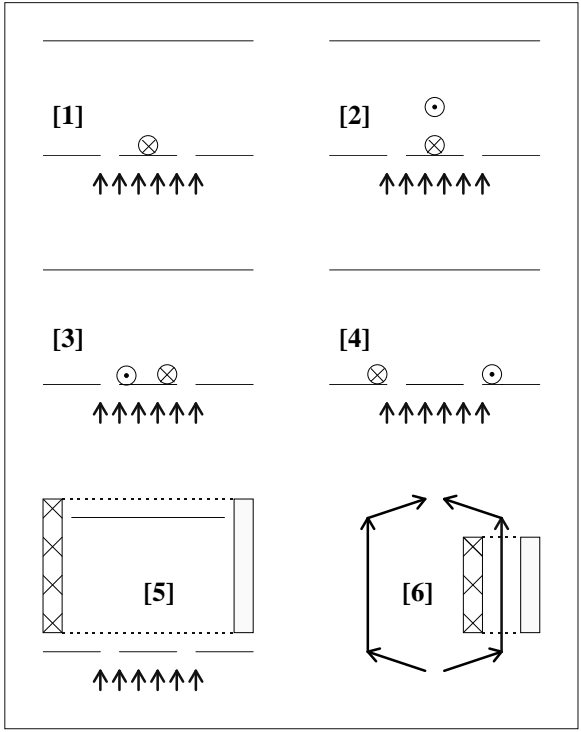
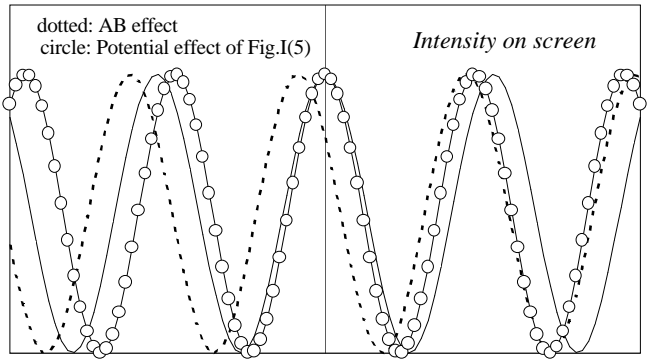
\end{document}